\documentclass[twocolumn,aps,showpacs,preprintnumbers,lettersize]{revtex4}
\usepackage{amsfonts}
\usepackage{amsmath}
\usepackage{amssymb}
\usepackage{graphicx}

\begin{document}

\title{Pseudogap induced by short-range spin correlations in a doped
       Mott insulator}
\author{B. Kyung, S. S. Kancharla, D. S\'{e}n\'{e}chal, A. -M. S. Tremblay}
\affiliation{D\'{e}partement de physique and Regroupement qu\'{e}b\'{e}cois sur les mat\'{e}riaux de pointe, Universit\'{e} de Sherbrooke, Sherbrooke, Qu\'{e}bec
J1K 2R1, Canada}
\author{M. Civelli, and G. Kotliar}
\affiliation{Physics Department and Center for Materials Theory, Rutgers University,
Piscataway, New Jersey 08855, USA}
\date{\today }

\begin{abstract}
We study the evolution of a Mott-Hubbard insulator into a correlated metal
upon doping in the two-dimensional Hubbard model using the Cellular 
Dynamical Mean Field Theory.
Short-range spin correlations create two additional bands apart from
the familiar Hubbard bands in the spectral function.
Even a tiny doping into this insulator causes a jump of the Fermi energy 
to one of these additional bands and an immediate momentum dependent
suppression of the spectral weight at this Fermi energy.
The pseudogap is closely tied to the existence of these bands.
This suggests a strong-coupling mechanism that arises from short-range 
spin correlations and large scattering rates
for the pseudogap phenomenon seen in several cuprates.
\end{abstract}

\maketitle


The issue of the origin of the pseudogap phenomenon observed in underdoped 
cuprates lies at the center of any theoretical 
explanation for high temperature superconductivity in the cuprates
and is one of the most challenging questions 
in condensed matter physics.
The suppression of low energy spectral weight in the normal state of these
materials 
has been observed through various experimental probes~\cite{TS:1999}.
In spite of many theoretical works to explain the observed anomalies,
there is no consensus at present.
The lack of
controlled approximations to deal with the strong coupling physics and
low dimensionality inherent to these systems continues to pose major
stumbling blocks towards a complete theoretical understanding.
Since the parent compounds of the cuprates are Mott-Hubbard insulators,
an understanding of such an insulator and its evolution into a correlated
metal upon doping is crucial.

In this paper we study the two-dimensional Hubbard model on a square lattice
at and near half-filling with Cellular Dynamical
Mean-Field Theory (CDMFT)~\cite{KSPB:2001}. 
The CDMFT method is a natural generalization of the single site DMFT~\cite{GKKR:1996} to incorporate short-range spatial correlations.
Since at and near half-filling short-range spin correlations are dominant at low energy,
this method is expected to describe
additional features caused by spin degrees of freedom
in the single-particle spectrum.
The CDMFT~\cite{BK:2002} has already passed several tests against exact results obtained by the Bethe Ansatz and the Density Matrix Renormalization Group (DMRG) techniques in one dimension, where the CDMFT scheme is expected to be in the worst case scenario.
Long-range order involving several lattice sites such as $d$-wave superconductivity can be also described in CDMFT~\cite{KCCKSKT:2005}. 
Several other cluster schemes have been
proposed~\cite{MJPH:2004,HTJPK:1998,SPP:2000,P:2003,LK:2000} 
including Dynamical Cluster Approximation (DCA)~\cite{HTJPK:1998},
Cluster Perturbation Theory (CPT)~\cite{SPP:2000} and its variational extension (V-CPT)~\cite{P:2003}.
The variational principle used in the last scheme allows one to consider CPT, V-CPT, and CDMFT within a unified framework. 

In the CDMFT construction~\cite{KSPB:2001,BK:2002} the infinite lattice is tiled with identical clusters of size $N_{c}$.
In an effective action description, the degrees of freedom in a single cluster are treated exactly, while the remaining ones are replaced by a bath of non-interacting electrons, which hybridizes with the cluster degrees of freedom.
For practical purposes, it is useful to view this cluster action as arising from a cluster-bath Hamiltonian of the form
\begin{eqnarray}
H &&=\sum_{\langle \mu \nu \rangle ,\sigma }t_{\mu \nu }c_{\mu \sigma
}^{\dagger }c_{\nu \sigma }+U\sum_{\mu }n_{\mu \uparrow }n_{\mu \downarrow }
\nonumber \\
+&&\sum_{m,\sigma }\varepsilon _{m\sigma }a_{m\sigma }^{\dagger }a_{m\sigma
}+\sum_{m,\mu ,\sigma }V_{m\mu \sigma }(a_{m\sigma }^{\dagger }c_{\mu \sigma
}+\mathrm{H.c.}) \, .  \label{eq20}
\end{eqnarray}%
Here the indices $\mu ,\nu =1,\cdots ,N_{c}$ label sites within the cluster, 
$m=1,\cdots ,N_{b}$ with $N_{b}$ representing the number of bath degrees of freedom, and $c_{\mu \sigma }$ and $a_{m\sigma }$ annihilate electrons on the cluster and the bath, respectively.
$t_{\mu \nu }$ is the hopping matrix within the cluster,
$\varepsilon _{m\sigma }$ is the bath energy and $V_{m\mu\sigma }$ is the bath-cluster hybridization matrix.
Let us adopt a matrix notation (cluster indices suppressed) for the hopping matrix $t$, the cluster Green function $G_c(i\omega)$, its noninteracting counterpart $G_0(i\omega)$ and the cluster self-energy $\Sigma = G_0^{-1}-G_c^{-1}$. 
We start with an initial guess for the bath parameters
$\varepsilon_{m\sigma}$ and $V_{m\mu\sigma}$ 
which determines a starting
$G_0(i\omega)$. With this guess the cluster Green function is
calculated by solving the cluster-bath Hamiltonian
(Eq.~\ref{eq20}). To close the self-consistency loop 
we obtain a new $G_0^{\prime}(i\omega)$ using
\begin{eqnarray}
G_{0}^{\prime -1}(i\omega _{n}) &=&\left( \frac{N_{c}}{(2\pi )^{2}}\int d\tilde{{\bf k}}%
\frac{1}{i\omega _{n}+\mu -t(\tilde{{\bf k}})-\Sigma (i\omega _{n})}\right) ^{-1}%
\kern-1em  \notag \\
&+&\Sigma (i\omega _{n})\;.  \label{eq41}
\end{eqnarray}
%
The integral comes from projecting on the cluster the Green function
of an infinite lattice formed of identical clusters located at positions
$\vec{R}_{i}$. The self-energy of that Green function is equal to that in a 
single cluster. In real space it vanishes for sites that do not belong 
to the same cluster. However, hopping between clusters is allowed in the 
same way as on the original infinite lattice. Hence, the hopping matrix 
has a dependence on superlattice vectors $\vec{R}_{i}-\vec{R}_{j}$ and 
correspondingly on wave vectors $\tilde{{\bf k}}$ that diagonalize that 
part of the problem.
Now to invert the relation between this $G_0^{\prime}(i\omega)$ and the bath parameters
for the next iteration we minimize 
\begin{eqnarray}
d = \sum_{\omega_n,\mu,\nu} \left| \left(
G_{0}^{\prime -1}(i\omega_n) - G_{0}^{-1}(i\omega_n) \right)_{\mu \nu} \right|^{2} \: .
\label{eq42}
\end{eqnarray}
The lattice Green function $G({\bf k},i\omega _{n})$ is obtained from 
\begin{eqnarray}
G({\bf k},i\omega _{n}) &=&\frac{1}{N_{c}}\sum_{\mu \nu }e^{i{\bf k}\cdot (%
\vec{r}_{\mu }-\vec{r}_{\nu })}  \notag \\
&\times &\left[ \frac{1}{i\omega _{n}+\mu -t(\tilde{{\bf k}})-\Sigma (i\omega _{n})}%
\right] _{\mu \nu } \;,  \label{eq43}
\end{eqnarray}
where $\Sigma (i\omega _{n})$ is the converged cluster self-energy,
${\bf k}$ is any vector in the original Brillouin zone and $\mu, \nu $ label cluster sites.
All quantities plotted in this paper are derived from this lattice Green function. 
This is different from the periodization of the self-energy
in Refs.~\cite{KSPB:2001,BK:2002,CCKPK:2005}.

\begin{figure}[b]
\includegraphics[width=8.0cm]{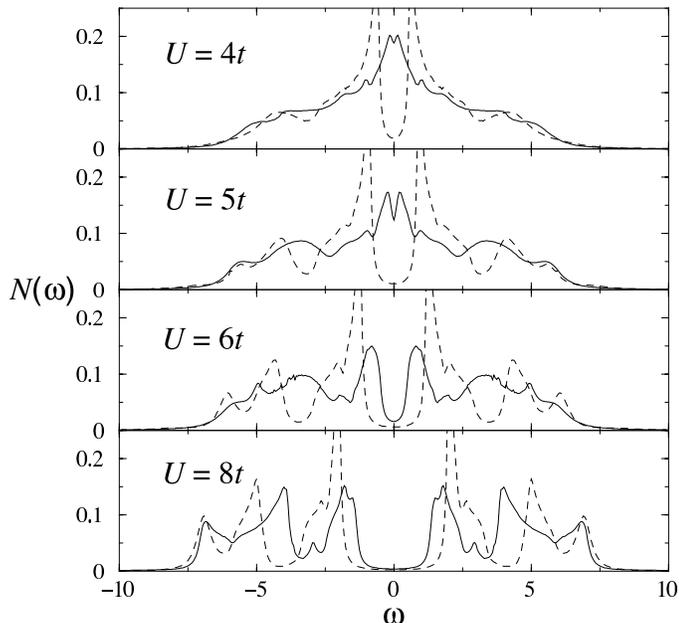}
\caption{Density of states $N(\protect\omega )$ at half-filling ($n=1$) for several values of $U$.
The dashed curves are $N(\protect\omega )$ with AF long-range order. }
\label{N_w_n=1_U.fig}
\end{figure}

Since spectra in and near a Mott insulator are expected to be highly singular, 
we used the exact diagonalization method~\cite{CK:1994}
to solve the cluster-bath Hamiltonian (Eq.~\ref{eq20}) at zero temperature.
This has the advantages of computing dynamical quantities directly 
in real frequency
and of treating the large $U$ regime without difficulty.
The self-consistency condition (Eq.~\ref{eq41}) was imposed on the imaginary frequency axis with a cutoff frequency larger than all the bath energies. 
In the present study we used $N_{c}=4$ sites for the cluster (minimum number of sites reflecting the full square lattice symmetry) and $N_{b}=8$ sites for the bath.  
Four-fold symmetry of the cluster leads to four independent bath parameters.
All energies are measured in units of $t$.
Results are plotted with a broadening parameter of $0.1t$.

\begin{figure*}[tbp]
\includegraphics[width=14cm]{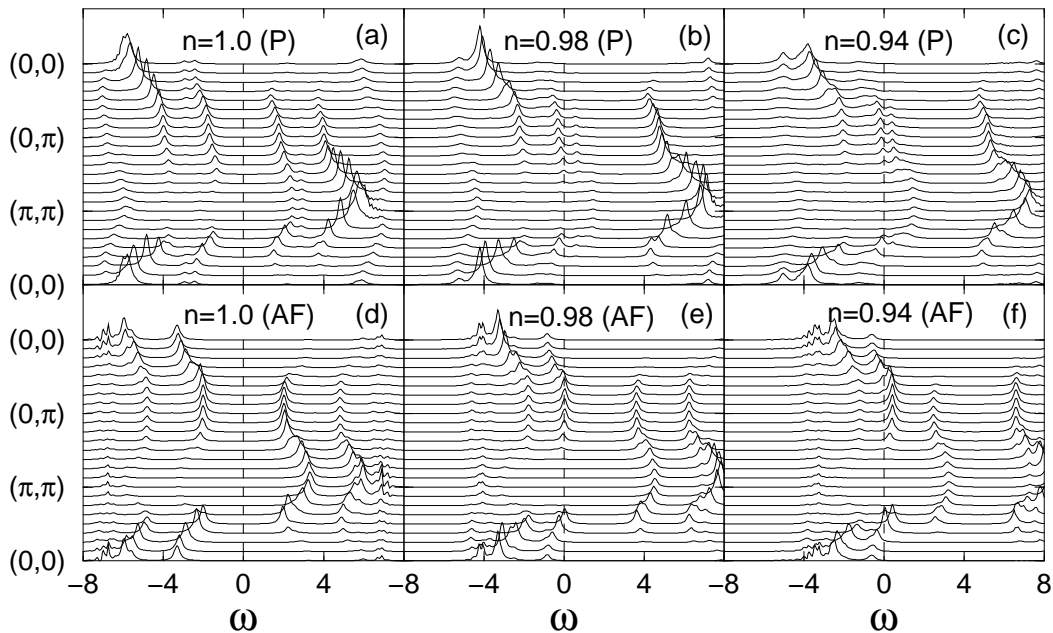}
\caption{Spectral function $A(\vec{k},\omega)$ along some symmetry 
         directions in the paramagnetic (upper row) and the AF   
         (lower row) states for several small dopings with $U=8t$.
         The dashed lines represent the Fermi energy.
        }
\label{A_k_w_n=1_0.98.fig}
\end{figure*}

In order to see how the Mott transition occurs in the presence of short-range correlations and how the differences between short- and long-range correlation effects evolve with interaction strength, we present in Fig.~\ref{N_w_n=1_U.fig} the density of states $N(\omega )$ as computed from the paramagnetic (P) (solid curves) and the antiferromagnetic (AF) (dashed curves) states at half-filling ($n=1$).
The two states (P and AF) are selected by imposing appropriate constraints on the bath parameters.
When the correlation length is larger than the size of a cluster, long-range effects in the single-particle spectrum are expected to be well described by the solution with the corresponding long-range order.
The difference between the two solutions at low energies is largest at weak coupling ($U=4t,5t$) and decreases at strong coupling ($U\gtrsim 8t$).
In the weak coupling case, $N(\omega )$ has only a small dip at the Fermi energy compared to the AF solution, indicating that short-range correlations available in a cluster of size $N_{c}=4$ are not long enough to lead to an insulating gap~\cite{MJ:2001,KLPT:2003}. 
As $U$ is increased to $6t$, two sharp bands begin to develop at the gap edge and become well separated from the Hubbard bands centered around $\pm U/2$. 
At this strength of $U$, the two solutions (P and AF) start approaching each other at low energies. 
When $U=8t$, an insulating gap is well defined between the low energy inner bands, and long-range correlations (dashed curve) sharpen these bands but they stay essentially at the same energy. Thus short-range correlations begin to dominate the low energy Physics at intermediate to strong coupling. 
At this value of $U$ the low energy inner bands are clearly distinguished from the Hubbard bands coming from purely local correlations. 
Since these bands are missed in the single site DMFT~\cite{Comment:DMFT} and smoothly evolve into sharp spin density wave-like bands when AF spin order is allowed, they are attributed to \textit{short-range spin} correlations. 
When $U$ is further increased beyond the bandwidth of $8t$ (not shown), the size of the insulating gap increases, but the relative shape and position of the inner and the Hubbard bands are unchanged.

   Next we investigate the evolution of a Mott-Hubbard insulator 
into a correlated metal upon a small electron doping at $U=8t$.
It is clear at half-filling 
(Fig.~\ref{A_k_w_n=1_0.98.fig}(a)) that 
$A(\vec{k},\omega)$ (and also $N(\omega)$ in Fig.~\ref{N_w_n=1_U.fig})
shows four significant features near $(0,\pi)$ and $(\pi/2,\pi/2)$.
The four peak structure in $A(\vec{k},\omega )$ was previously found by Moreo \textit{et al.}~\cite{MHSD:1995} and Preuss 
\textit{et al.}~\cite{PLH:1995} in their high resolution Quantum Monte Carlo (QMC) results for the half-filled 2D Hubbard model.
As will be shown shortly, the existence of the additional bands 
apart from the Hubbard ones turns out to be crucial for the systematic 
physical picture of the pseudogap phenomenon.

   The most striking feature of even a tiny doping into a Mott-Hubbard
insulator (Fig.~\ref{A_k_w_n=1_0.98.fig}(b) and (c)) is that one of the inner
bands at $\omega > 0$ seems to disappear immediately while the Fermi energy jumps to the
other band that persists at $\omega < 0$. Some of the weight is transferred
near the Fermi energy at $\omega > 0$~\cite{EMS:1991}.
Furthermore, one already sees here the pseudogap phenomenon.
Especially for $n=0.94$, one clearly sees that
the peak dispersing from low binding energy near $(\pi/2,\pi/2)$ becomes sharper close to the
Fermi energy while near $(\pi,0)$ the corresponding peak becomes broader
and less intense. The Fermi energy tends to sit towards a minimum instead
of a maximum near that point.
The suppression of weight near the Fermi energy eventually disappears at about $20\%$ doping.
The appearance of a low energy peak together with one at high binding energy
is consistent with exact diagonalization 
results for the $t-J$ model~\cite{MHSD:1995,Dagotto:1994}.
We also show clear differences between the momentum resolved spectra of the paramagnetic and the AF states.
The main differences for the AF case at half-filling (Fig.~\ref{A_k_w_n=1_0.98.fig}(d)) are that the low energy bands are more intense than in the paramagnetic state (Fig.~\ref{A_k_w_n=1_0.98.fig}(a)) and that they bend back near $\left( 0, \pi\right) $ and near $\left( \pi /2,\pi /2\right)$, reflecting the underlying AF symmetry.
The results of Fig.~\ref{A_k_w_n=1_0.98.fig}(d) are strikingly similar to those obtained in a 10 site V-CPT calculation~\cite{Dahnken03}. 
The more drastic differences between the paramagnetic and AF solutions occur at finite doping.
While in both cases the Fermi level immediately jumps to the closest low energy band upon doping, Figs. \ref{A_k_w_n=1_0.98.fig}(e) and (f)~\cite{Comment:AF} show that, in the AF solution, the four peak structure is preserved and the excitations at the Fermi energy are quite sharp.

\begin{figure}[tbp]
\includegraphics[width=8.5cm]{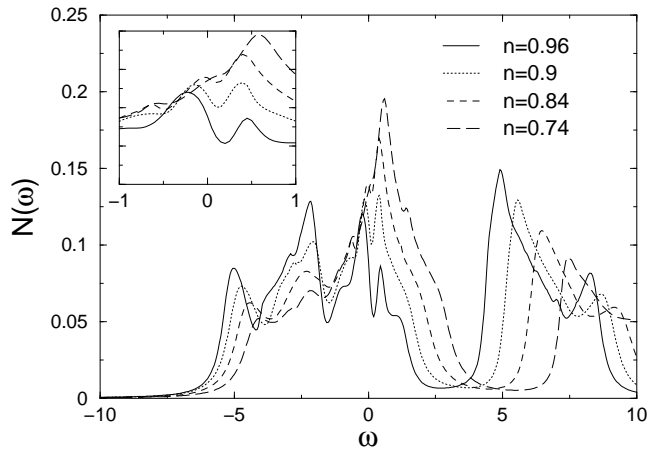}
\caption{Density of states $N(\protect\omega)$ for several doping levels with $U=8t$.
The solid, dotted, dashed, and long-dashed curves correspond to 
$n=0.96$, $0.9$, $0.84$, and $0.74$, respectively.
The inset shows $N(\protect\omega)$ near the Fermi energy.}
\label{N_w_pseudogap_density.fig}
\end{figure}

Fig.~\ref{N_w_pseudogap_density.fig} shows the density of states $N(\omega)$ for several dopings.
The pseudogap found here (inset) occurs over a scale $0.3t-0.4t$  
and vanishes for large enough doping in agreement  
with previous works~\cite{Huscroft:2001,SP:2003}. 
However, we have checked that at large $U$ the
scale of the pseudogap,  
as defined by the distance between the peaks closest to zero in 
Fig.~\ref{N_w_pseudogap_density.fig},
is almost independent of $U$ in agreement with
Ref.~\cite{ST:2004}. It does not scale as $J$, contrary to Ref.~\cite{SP:2003}.
Furthermore, the pseudogap size increases with decreasing doping, consistent with several recent experiments in the cuprates~\cite{TS:1999,DHS:2003}.
The pseudogap feature disappears not because the two peaks coalesce but because of a gradual filling up of the spectral weight between the two peaks. 
Note the evolution with doping of the asymmetry about the Fermi energy.
It is qualitatively consistent with what is expected for a doped Mott 
insulator~\cite{EMS:1991}.
We also wish to point out clear evidence of correlation
between the existence of pseudogap at the Fermi level (inset) 
and the separation of the low energy bands from the Hubbard ones.
The appearance of a pseudogap near $n=0.8$ found in the single-particle
spectrum coincides with the downturn of the uniform cluster spin susceptibility 
(not shown), suggesting that the pseudogap has to do with 
spin singlet formation of correlated electrons.

\begin{figure}[tbp]
\includegraphics[width=8.5cm]{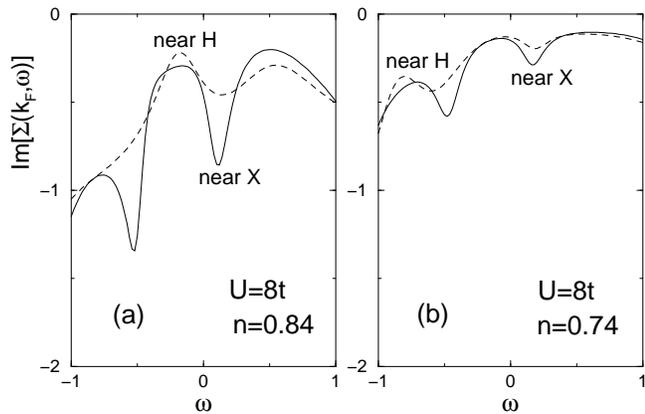}
\caption{Imaginary part of the self-energy for 
    (a) $n=0.84$ at $\vec{k}=(   0  , 52/64 )/\pi$ (solid line) 
                and $        ( 30/64, 30/64 )/\pi$ (dashed line),
and (b) $n=0.74$ at $\vec{k}=(   0  , 49/64 )/\pi$ (solid line) 
                and $        ( 29/64, 29/64 )/\pi$ (dashed line).
X$=(0,\protect\pi )$ and H$=(\protect\pi /2,\protect\pi /2)$.}
\label{self_energy.fig}
\end{figure}

The significant differences of spectral properties at $n=0.84$ and $n=0.74$
(Fig.~\ref{N_w_pseudogap_density.fig}) can be understood in terms of self-energies~\cite{ST:2004}. Fig.~\ref{self_energy.fig}(a) shows the imaginary part of the self-energy for $n=0.84$ at the Fermi wave vectors~\cite{Comment:Fermi_wave_vector}
near $(0,\pi )$ and $(\pi /2,\pi /2)$.
The scattering rate becomes large near X in particular, leading to the pseudogap in the single-particle spectral function.
However, Fig.~\ref{self_energy.fig}(b) shows much smaller self-energy in magnitude for $n=0.74$, leading towards the Fermi liquid at large doping.
Clearly, upon doping a Mott insulator into a metallic state, quasiparticles disappear since the absolute value of the imaginary part of the self-energy is a local maximum at the
Fermi energy instead of minimum as 
in a Fermi liquid picture (see caveat \cite{Comment:Resolution}).
The large scattering rate leading to the pseudogap found here is reminiscent of that due to the presence of the large density of states for scattering at the van Hove singularity in the weak coupling case.
With a finite second-neighbor hopping $t^{\prime}$, the crossing of the non-interacting 
Fermi surface and the AF Brillouin zone boundary leads to additional  
scattering processes~\cite{ST:2004}. We now discuss this case.

%

\begin{figure}
\hspace{-1.6cm}
\resizebox{10.0cm}{!}{\includegraphics[angle=270]{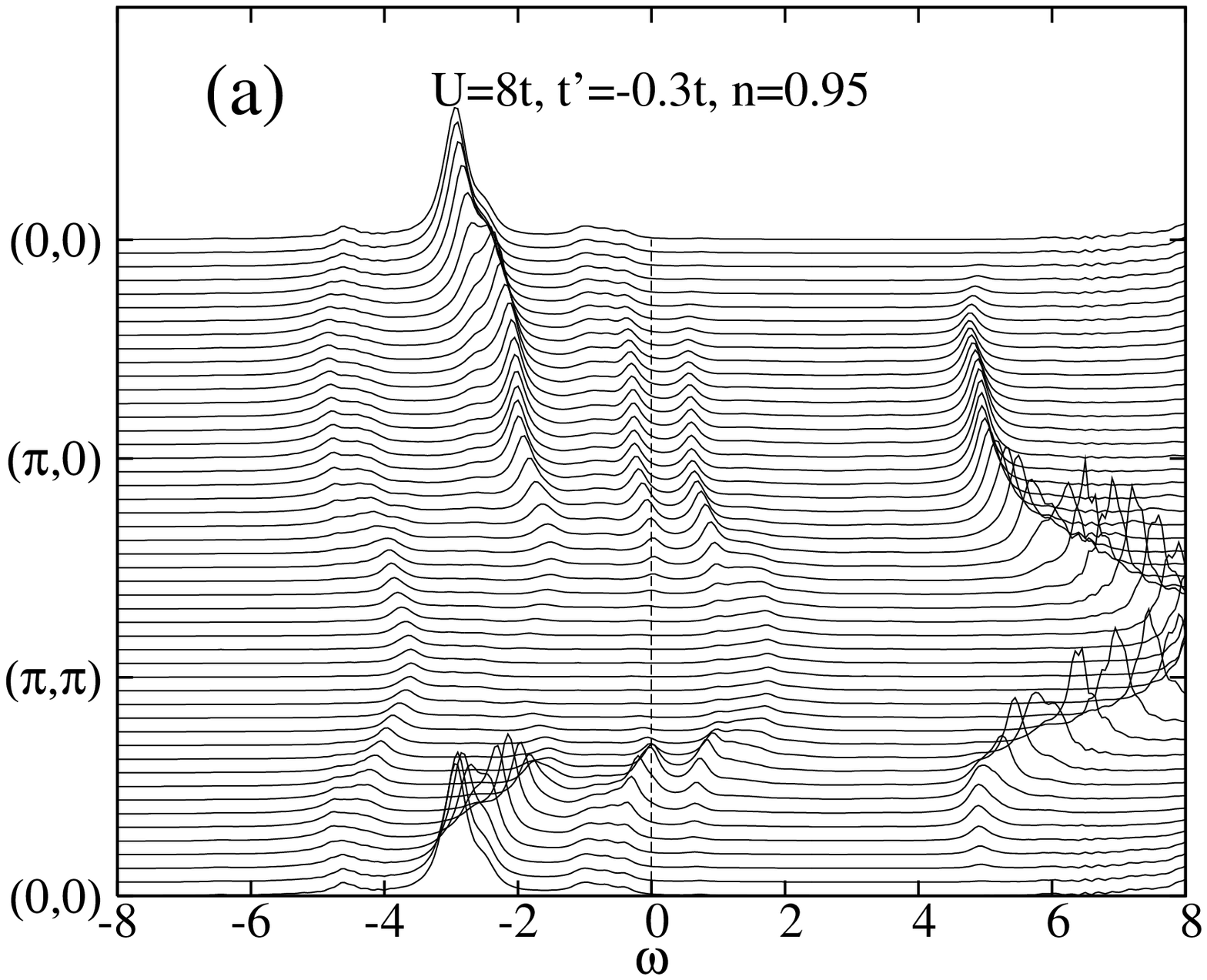}}
\par\vspace{0.5cm}
\hspace{-1.6cm}
\resizebox{10.0cm}{!}{\includegraphics[angle=270]{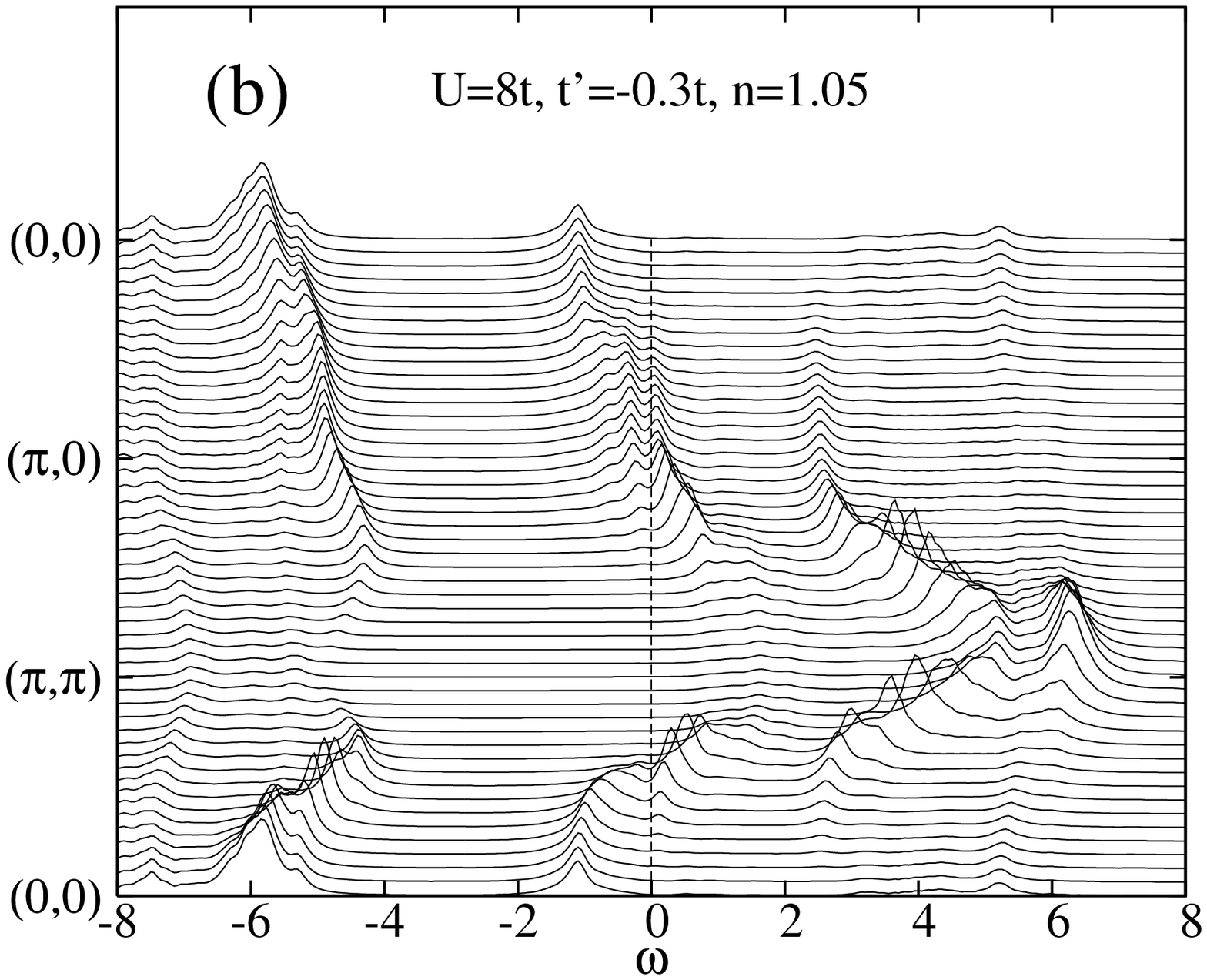}}
\caption{Spectral function $A(\vec{k},\omega)$ along some symmetry 
         directions in the $t-t^{\prime}-U$ model with
         $t^{\prime}/t = -0.3$ for $5 \%$  
         (a) hole- and (b) electron-doping. 
         The dashed lines represent the Fermi energy.
        }
\label{A_k_w_8.0_2.09_-.30_fine.fig}
\end{figure}

\begin{figure}
\resizebox{4.0cm}{!}{\includegraphics{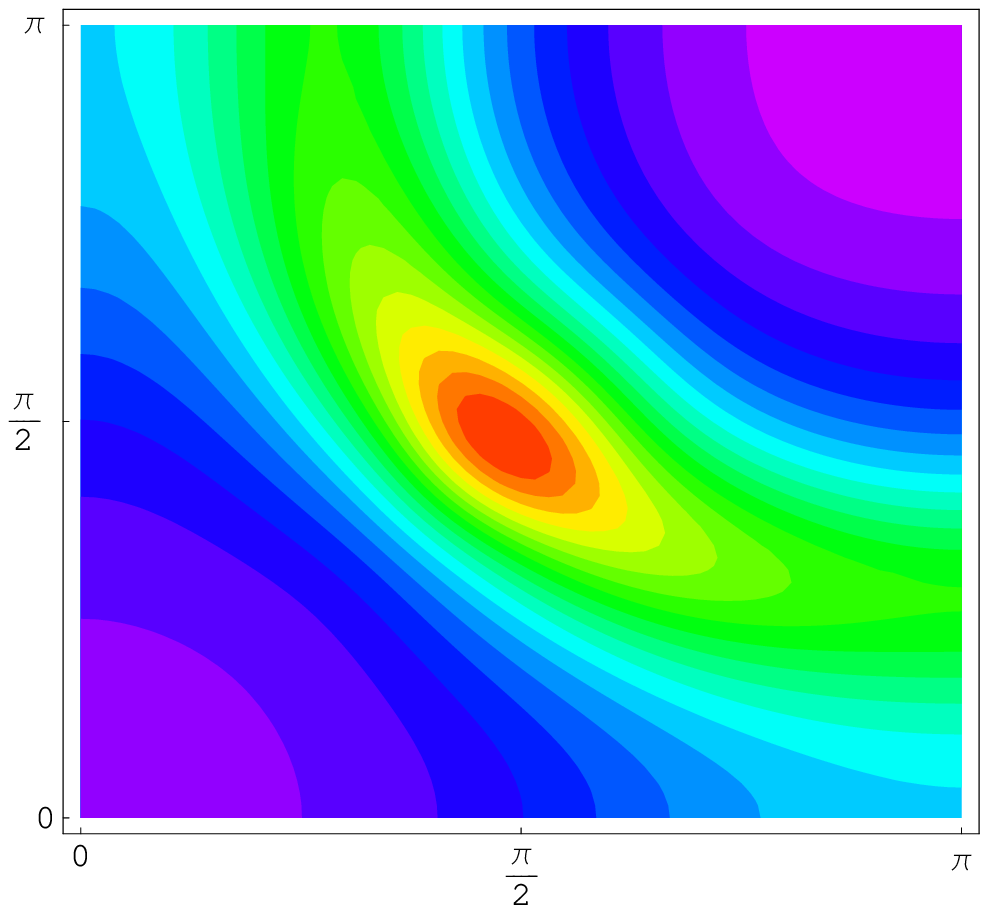}}
\includegraphics[width=4.0cm]{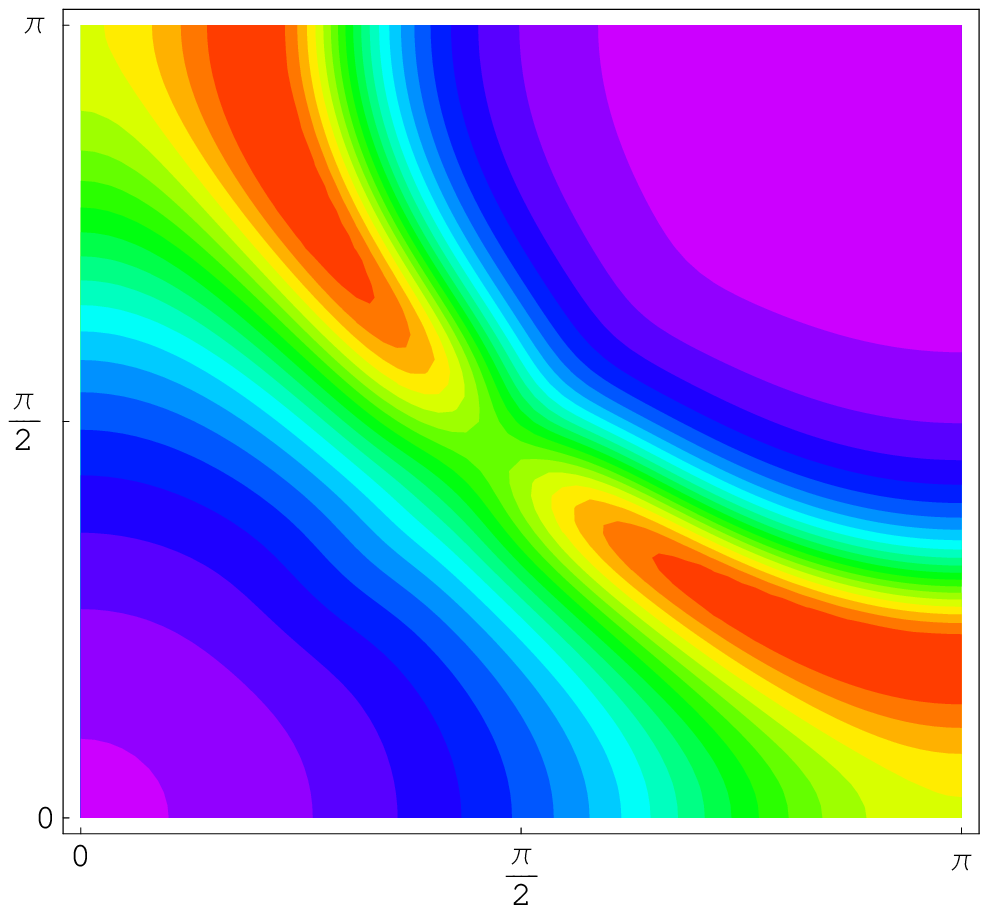}
\caption{Density plot $A(\vec{k},0)$ in the $t-t^{\prime}-U$ model with
         $U=8t$, $t^{\prime}/t = -0.3$ for $5 \%$
         hole- (left) and electron-doping (right),
         i.e. $n=0.95$ and $n=1.05$, respectively.
         The current plots were obtained by taking the average of 
         $A(\vec{k},\omega)$ in a $(-0.2t,0.2t)$ window around 
         the Fermi energy.
        }
\label{A_0_8.0_2.09_-.30.fig}
\end{figure}

   We study the asymmetry of hole- and electron-doped cuprates  
in their spectral function in the $t-t^{\prime}-U$ model with 
$t^{\prime}/t = -0.3$~\cite{Macridin:2005} for $5 \%$ hole- and electron-doping.
As in the case of $t^{\prime} = 0$, the pseudogap shown  
in Fig.~\ref{A_k_w_8.0_2.09_-.30_fine.fig} appears in 
the low energy band caused by short-range spin correlations, which is 
well separated from the Hubbard bands persisting at high energies.
However, the locations where the pseudogap occurs are opposite 
for hole- and electron-doping.
For hole-doping, $A(\vec{k},\omega)$ is gapped near $\vec{k}=(\pi,0)$,
while for electron-doping it is gapped near $\vec{k}=(\pi/2,\pi/2)$,
in agreement with ARPES 
experiments~\cite{Armitage:2001,Armitage:2002,Ronning:2003,Matsui:2005}.
This is clearer on the corresponding density plots in 
Fig.~\ref{A_0_8.0_2.09_-.30.fig} that show the asymmetry of
hole- and electron-doping at $\omega=0$ consistent with ARPES experiments.
The combination of the above two plots indicate that the pseudogap 
is associated not only with the suppression of the spectral weight, 
but also with its shift from the Fermi level.
Since the spectral function $A(\vec{k},\omega)$ is obtained on a 
small cluster in the paramagnetic state, short-range spin correlations 
suffice for the pseudogap at both dopings.
This is in agreement with CPT results~\cite{ST:2004} and exact diagonalizations on the $t-t^{\prime}-t^{\prime\prime}-J$ model~\cite{Tohyama:2004}, but in disagreement with 
recent DCA results~\cite{MJMK:2005} where AF long-range order (or a large AF correlation length) is needed 
for a pseudogap under electron doping.

   It is remarkable that including only the nearest-neighbor correlations into this two-dimensional model is enough to capture some of the details of
the spectral properties that are missed in the single site DMFT. 
We stress that at 
\textit{weak} coupling and in two dimensions, a different mechanism involving large AF correlation lengths causes the pseudogap~\cite{KLPT:2003,ST:2004,KHDT:2004,SNS:2004}.
That mechanism, relevant for electron-doped cuprates near optimal doping, is clearly not involved in the present \textit{strong} coupling calculations on small clusters.
Although the sizes of clusters accessible to computation limit the resolution of the spectral properties, we expect the above picture for the \textit{strong} coupling pseudogap to be robust with respect to an increase in the cluster size. 
This is confirmed by 
our recent low (but finite) temperature CDMFT+QMC calculations~\cite{KKT:2005}
where at intermediate to strong coupling
a $2 \times 2$ cluster accounts for more than $95 \%$ of the correlation
effect of the infinite size cluster in the single-particle spectrum.

  Our calculations were done at zero temperature, while the pseudogap
behavior is observed at finite temperature and broken-symmetry states 
appear in most of the zero temperature phase diagram.
Nonetheless, our present results would be most relevant when the temperature
is smaller than the characteristic energy scale for the pseudogap         
or in regions where broken-symmetries are absent or are destroyed at zero
temperature.
In fact the latter situation was precisely realized in a series of 
experiments~\cite{Boebinger:1996,Fournier:1998,Ono:2000} 
on the cuprates, which have shown a metal-insulator crossover
near optimal doping by suppressing the superconducting phase with
a high magnetic field, revealing the underlying pseudogap state at zero
temperature.
Recent experiments~\cite{Pimenov:2005} also show that when superconductivity
is suppressed by Ni doping, the pseudogap survives at very low 
temperature in the corresponding paramagnetic state.
Our results could be also relevant even in broken-symmetry 
states, as long as the sizes of the corresponding gaps are much smaller than   
the characteristic energy scale for the pseudogap.
Even if we work 
at $T=0$, the limited size of our clusters prevents long-range order
unless we add mean-fields. Hence our calculation mimics the normal state.

   In conclusion, the long standing problem of the evolution of a Mott-Hubbard insulator into a correlated metal upon doping has been examined in the two dimensional Hubbard model by using the Cellular Dynamical Mean Field Theory
(CDMFT) which incorporates short-range spatial correlations.
At half-filling these correlations create two additional bands besides the familiar Hubbard ones.
Even a tiny doping into a Mott-Hubbard insulator ($t^{\prime}=0$) causes the Fermi energy to jump to one of these bands and the spectral weight to be suppressed immediately in a ${\bf k}$ dependent way, such that a pseudogap appears near $\left( \pi ,0\right) $ while a peak survives near $\left( \pi
/2,\pi /2\right) $.
When AF long-range order is present at finite doping, the picture is quite different, demonstrating that at strong coupling the pseudogap in the particle-hole symmetric model is a short-range effect.
Including particle-hole asymmetry with a second neighbor hopping $t^{\prime}$
gives spectral weights that are similar to those observed in hole- and 
electron-doped cuprates.
This provides a systematic physical picture for the emergence of the strong coupling pseudogap phenomenon in doped Mott-Hubbard insulators and the consequent non-Fermi liquid behavior that arises solely from short-range spin correlations~\cite{Huscroft:2001,SP:2003,ST:2004} and large scattering rates, without any symmetry breaking.

We thank O. Parcollet for useful discussions.
The present work was supported by NSERC (Canada), FQRNT (Qu\'{e}bec), CFI
(Canada), CIAR, the Tier I Canada Research Chair Program (A.-M.S.T.), and the NSF under grant DMR-0096462 (G.K.).


\begin{thebibliography}{99}
\bibitem{Comment:Fermi_wave_vector} Since the Fermi surface itself is not 
well defined in our results, we computed the imaginary part of the
self-energy at the noninteracting Fermi wave vectors. The results are 
robust with modest variations in $\vec{k}$.
\bibitem{Comment:Resolution} Due to the finite bath and cluster sizes,
our calculations are limited to
an energy resolution of about $0.1t$. Hence the possibility of a Fermi
liquid regime at lower energies cannot be entirely ruled out. 
Nevertheless this energy resolution is much smaller than the characteristic 
energy scale for the pseudogap. The location of the maximum in the 
absolute value of the self-energy should also not be more accurate than 
$0.1t$.
\end{thebibliography}
\end{document}